\documentclass[twocolumn]{jpsj2}
\usepackage{bbm}

\title{Magnetic Excitations in Bilayer
High-Temperature Superconductors  with Stripe Correlations}

\author{G.S.\ \textsc{Uhrig}$^{1}$\thanks{E-mail address:
{gu@thp.uni-koeln.de}}, K.P.\ \textsc{Schmidt}$^{1}$
 and M.\ \textsc{Gr\"{u}ninger}$^{2}$}

\inst{$^{1}$Institut f\"{u}r Theoretische Physik, Z\"{u}lpicher Stra\ss{}e 77,
Universit\"{a}t zu K\"{o}ln, 50937 K\"{o}ln, Germany\\
$^{2}$2. Physikalisches Institut, RWTH Aachen, 52056 Aachen, Germany}

\abst{
The universal behaviour of the magnetic excitations in
high-temperature superconductors is described in a model with static
stripes retaining only the localized spin degrees of freedom.
The stripes are represented by a model of coupled two-leg spin ladders.
We start from the results obtained previously by continuous unitary
transformations for an isolated spin ladder.
A quantitative description of neutron scattering data is reached, using
a model for a single cuprate layer with well established values of the
exchange coupling constants. The neutron resonance peak is explained in
terms of a saddle point in the dispersion of the magnetic excitations.
Here we make predictions for bilayer systems with in-phase or out-of-phase
stripe correlations.
The results may serve as a guide for future experimental analyses.
}

\kword{magnetic excitations, spectral densities, stripe phase,
inelastic neutron scattering, high-temperature superconductors}

\begin{document}
\maketitle

\section{Introduction}
The r\^ole of the magnetic excitations in the mechanism of
high-temperature superconductivity is still heavily debated.
A prerequisite for a successful understanding of this r\^ole
is a quantitative description of the magnetic excitations
themselves. For many years, however, it seemed that different
families of compounds display rather different behaviour
of their magnetic excitations \cite{brom03,norma03,sidis04}.
Recent experimental evidence for a universal behaviour of the magnetic
excitations \cite{hayde04,tranq04a,chris05}
clearly indicates the relevance of these features and calls for
detailed theoretical investigations.

\paragraph{Phenomena}
In the single-layer compound La$_{2-x}$Sr$_x$CuO$_4$ (LSCO),
four incommensurate satellites
are observed in elastic neutron scattering experiments
at low temperatures. These satellites are shifted away
from the antiferromagnetic wave vector
${\bf Q}_{\rm AF} =(1/2,1/2)$ in reciprocal lattice units (rlu).
In the insulating phase at low doping ($p \lessapprox 6\%$),
the shift occurs along the diagonal while it occurs parallel to
the tetragonal reciprocal axes in the superconducting phase
($p \gtrapprox 6\%$) \cite{wakim99,fujit02}.
The size of the shift is roughly linear in the doping $p$ for
small values of $p$. It saturates at about 1/8 of the
Brillouin zone for sizeable doping levels ($p \gtrapprox 10\%$)
\cite{yamad98,fujit02}, see also Fig.\ 1.4 in Ref.\ \cite{castr03}.
These incommensurate features may be explained in terms of the superstructure
satellites of a stripe phase (see below),
which is supported by the observation of the
corresponding charge-order satellites around the Bragg peaks in Nd-doped LSCO
\cite{tranq95,zimme98}.
Incommensurate magnetic excitations were observed in LSCO also in
{\em inelastic} neutron scattering (INS) experiments
\cite{cheon91,mason96,aeppl97,lake99,hirak01,tranq04b,chris05}.

In the bilayer compound YBa$_2$Cu$_3$O$_{x}$ (YBCO$_x$),
the magnetic response at optimum doping is dominated below $T_c$
by the resonance peak at ${\bf Q}_{\rm AF}$
with an energy of $\omega_{\rm res} \approx 41 $ meV
\cite{rossa91,brom03,sidis04}.
The observation of a resonance peak at ${\bf Q}_{\rm AF}$ below $T_c$
with $\omega_{\rm res} \approx 47$ meV in single layer
Tl$_2$Ba$_2$CuO$_{y}$
\cite{he02} shows that the bilayer structure is not a prerequisite for the
occurrence of the resonance peak.
Incommensurate branches have been observed at low energies in the underdoped
regime  and more recently also at about optimal doping
\cite{arai99,bourg00a,rezni04,yokoo03a,yokoo03b}.
Elastic superstructure satellites indicating charge order were also reported in
underdoped YBCO \cite{mook02a}, suggesting that
also this system may display an instability towards stripe formation.
A third feature is the appearance of incommensurate branches
{\em above} the resonance
energy \cite{arai99,mook02a,rezni04,yokoo03a,yokoo03b}.

\paragraph{Universality}
The data for YBCO provide evidence that the resonance peak and
the incommensurate excitations are not separate features but form part of the
common dispersive magnetic excitations of the cuprate high-temperature
superconductors. Very recently, this point of view has been
substantiated by reports on very similar INS data
obtained on YBCO$_{6.6}$ \cite{hayde04} and
La$_{15/8}$Ba$_{1/8}$CuO$_4$ (LBCO)
\cite{tranq04a}, both at low and at high energies in a large part of the
Brillouin zone. On the one hand, the experiment of Tranquada and co-workers
on LBCO \cite{tranq04a} provides data for a \emph{charge-ordered} phase which
suppresses the superconducting phase \cite{moode88,fujit04}.
The data show the incommensurate excitations which are familiar in the
La family. Additionally, a dominant feature is observed at ${\bf Q}_{\rm AF}$,
which is the first report of such a resonance in this family.
In contrast to YBCO, the resonance in LBCO is observed in the normal
(charge-ordered) state. On the other hand, the experiment by Hayden and
co-workers for YBCO$_{6.6}$ displays very similar features \cite{hayde04}.
Yet the system is in the \emph{superconducting} state.
Recently, Christensen {\em et al.} pointed out the similarity of the dispersion
in the superconducting phases of optimally doped LSCO and YBCO \cite{chris05}.

These findings suggest that the magnetic excitations
in the high-$T_c$ cuprate superconductors are of universal character.
Differences may stem from the differences in the state of the charges.
The absence of a resonance peak in LSCO may be explained by the position of the
resonance mode with respect to the particle-hole continuum.
If the decay into particle-hole pairs is fast the magnetic modes may be
overdamped so that they are not seen in experiment as prominent peaks
\cite{morr98,abano99}.
In fact, the incommensurate magnetic excitations in LSCO become much sharper
upon cooling \cite{mason96,aeppl97}.

\paragraph{Modulated Phases}
An appealing route to account for the important
features of the magnetic excitations
sketched above is to assume a certain long-range charge modulation.
This implies a corresponding superstructure
in the spin sector which in turn leads to
periodicities different from those of the underlying lattice.
The most prominent modulation considered is the formation of stripes.
It was proposed theoretically in the late eighties \cite{schul89,zaane89}
on the basis of the energy gain due to binding of charges to domain walls.
Later such modulations were observed experimentally in, e.g.,
nickelates \cite{chen93,tranq94}, and cuprates \cite{tranq95}.

While there are certain cuprate systems for which the existence of static
stripes is well established,
e.g., rare-earth-doped LSCO \cite{tranq95,zimme98,klaus00},
there are others which do not appear to display static long-range
stripes, for instance optimally doped YBCO.
We like to emphasize, however, that there are theoretical predictions
based on phenomenological dimer models which show
that even very small charge modulations can have a sizeable effect
on the spin sector, e.g., inducing a significant modulation of dimer
correlations \cite{balen05b}.
So it is not excluded that  certain charge orderings eluded so far
experimental observation because of their smallness.

Recent experiments on an {\em untwinned} sample of slightly underdoped
YBCO$_{6.85}$ \cite{hinko04} have not found a significant anisotropy
in the weights of the incommensurate peaks. The data display a certain
anisotropy regarding the peak width. Assuming that a given orthorhombic
domain determines the orientation of a possible stripe order these
findings do not support the existence of a static
one-dimensional stripe pattern, but they may be consistent with fluctuating
stripes. For a further discussion of the physics of static or dynamic stripes
we refer to Refs.\ \cite{brom03,emery97,kivel03,castr03}.

\paragraph{Stripe Pattern}
The stripe phase corresponds to a segregation into hole-rich and hole-poor
ribbons. Experimentally, a doping
dependence of the periodicity has been reported in elastic neutron scattering
experiments on underdoped LSCO \cite{yamad98,fujit02}.
The shift of the magnetic satellites away from ${\bf Q}_{\rm AF}$ saturates
at about 1/8 for larger doping levels ($p \approx 0.1$), see e.g.\ Fig.\ 1.4
in Ref.\ \cite{castr03}, corresponding to a spin superstructure periodicity of
$8a$ where $a$ is the in-plane lattice constant.
The concomitant charge superstructure periodicity of static stripes
is found to be $4a$ \cite{tranq95,zimme98}.
This is the periodicity we choose in our model.
The spin superstructure of $8a$ can be explained by, e.g., assuming that
the hole-rich ribbons form anti-phase domain walls for the spins, corresponding
to an effective {\em ferromagnetic} coupling between the spins across a
hole-rich stripe (see below).

In the literature, two patterns of stripe modulation are studied.
One is the site-centered pattern where the hole-rich stripe
is a chain, i.e.\, the width of this stripe is only one site.
Neighbouring hole-rich stripes are separated by three hole-poor sites with
localized spins, which can be viewed as a three-leg spin ladder.
This pattern is the one that was considered mostly \cite{brom03}.
But an attractive alternative is the bond-centered pattern where
the hole-rich stripe has a width of two sites.
These stripes are separated by two hole-poor spin sites, corresponding
to a two-leg spin ladder (see Fig.\ \ref{fig:sketch}).
There is evidence from band structure calculations that the bond-centered
modulation is favourable  \cite{anisi04}.
Moreover, the magnetic state is more stable in the bond-centered pattern than
in the site-centered one because spin ladders
with an even number of legs are gapped \cite{schul86,dagot96}
and thus stable against small perturbations. Spin ladders with
odd number of legs are critical \cite{schul86,dagot96} and thus highly unstable
against any perturbation or spontaneous symmetry breaking.
A quantum Monte-Carlo study illustrates that in the bond-centered
pattern much of the fluctuations originate directly in the
spin sector while in the site-centered pattern the interplay
between spin and charge sector is essential \cite{tworz99}.

Another interesting piece of evidence in favour of the
bond-centered scenario can be derived from the Fourier analysis
of the modulations observed by scanning tunneling microscopy (STM)
\cite{hanag04} on underdoped Ca$_{2-x}$Na$_x$CuO$_2$Cl$_2$.
We interpret the data here as resulting from
the spatial average of vertical and horizontal stripes.
A discussion of a possible truly two-dimensional
$4\times 4$ tiling modulation is left to future research.
The STM data show features at $\pm 1/4$ rlu, but not at $1/2$ rlu.
The  bond-centered modulation with period $4a$
is indeed generated by a single cosine term
with wave vector $2\pi/4a$; no higher harmonics appear.
In contrast, the general site-centered modulation with period $4a$
is characterized by the fundamental and the first harmonic.
Hence, the generic site-centered modulation should display
a feature also at $1/2$ rlu.
Thus the STM data rather support the bond-centered scenario.

\paragraph{Theoretical Approaches}
There are many theoretical approaches to the magnetic excitations
in the cuprate superconductors \cite{norma03}.
They can be split into two classes:\\
(i) Starting from an underlying \emph{fermionic} model,
mostly extended Hubbard models, one has to deal with strong interactions.
The magnetic collective modes appear in the particle-hole
or particle-particle channel. In essence, they are bound states or
resonances of two fermions \cite{norma03}. Widely used techniques
in this field are approaches based on random phase approximations
(see e.g.\ Ref.\ \cite{schny04} and references therein),
and renormalization schemes (see e.g.\ Refs.\ \cite{zanch00,furuk00,halbo00a}).
The regime of strong interactions with coupling constants larger than
the band width is difficult to describe reliably.
\\
(ii) Alternatively, one may start from a bosonic model which contains the
collective modes already from the very beginning.
Then the interplay with the charge degrees of freedom is added by some
coupling.
The approaches based on  spin-fermion models \cite{morr98,abano03}
or many treatments of $t$-$J$ models \cite{schmi88,dagot94b} are of this type.
In the very limit, the fermionic excitations are neglected, focusing
on the collective modes alone (see e.g.\ Refs.\ \cite{demle04,balen05a}).

In the present work we use an approach of the second type.
We discuss a spin-only Heisenberg model of coupled spin ladders
with effective coupling parameters,
assuming that the charge degrees of freedom are integrated out.
Roughly speaking, the
excitation energies of the charge degrees of freedom are
of the order of $t$ while the magnetic energies are of the order of $J\approx
t/3$. So it makes sense to consider an effective magnetic model at low
energies with the faster charge degrees of freedom being eliminated. One has
to keep in mind that the decay of the magnetic collective modes into
particle-hole pairs is neglected which will certainly play a r\^ole at higher
energies. Neglecting this decay is well justified in the superconducting
state where the charge excitations are mostly gapped.

While we  start from a spin-only model with long-range stripe order
our predictions are equally relevant
for systems with fluctuating stripe order for not too low energies.
The time scale of these fluctuations should be above
the time scale set by the inverse energy of the features under study.

Vojta and Ulbricht \cite{vojta04} investigated the
same spin-only model without the cyclic spin exchange
(see below) by a mean-field approach starting from
 dimers on the rungs of the ladders. The description of isolated ladders is
improved by a local energy correction which accounts for
a part of the effect of the hard-core constraint of the excitations
on each dimer. The constant-energy scans in this approach agree qualitatively
with the results we obtained for a single-layer model
\cite{uhrig04a,uhrig05a}.

In a phenomenological approach Vojta and Sachdev considered
also a plaquette modulation \cite{vojta05a}. From their results they concluded
that the magnetic excitations of a plaquette modulation do not agree
with the inelastic neutron data \cite{tranq04a,hayde04}.

Seibold and Lorenzana performed a large-scale time-dependent Gutzwiller
calculation for a Hubbard model to determine the magnetic excitations of a
stripe phase with charge and spin order \cite{seibo05}. Using parameters
adapted to describe the magnetic dispersion of the
undoped system and the doping dependence of the
incommensurability they find good agreement with the inelastic neutron data
for  LBCO. The similariy between their constant-energy scans and those
from spin-only approaches suggests that the damping due to
the charge degrees of freedom does not play an essential r\^ole in the
presence of charge and spin order. We presume that this is due to a partial
freezing out of the charge degrees of freedom.

\paragraph{Aim of this Study and Set Up}

It is the aim of the present study to extend our previous
investigations based on long-range stripe modulations in a single layer
\cite{uhrig04a,uhrig05a} to bilayer cuprates such as YBCO.\@
In addition, further results for single layers will be presented.
All these results provide valuable guiding predictions for present and
future inelastic neutron scattering experiments.

The article is set up as follows. In the next section, Sect.\ \ref{sec:model},
the model will be introduced and its theoretical treatment will be
explained. In Sect.\ \ref{sec:monolayer} results for single layers
will be presented. In the subsequent Sect.\ \ref{sec:bilayer}, results
for two different modulation patterns in bilayers will be shown and compared.
In Sect.\ \ref{sec:scans} various scans at constant energy will be given.
Finally, the conclusions will be drawn in Sect.\ \ref{sec:conclusio}.

\section{Model and Calculation}
\label{sec:model}
\begin{figure}[t]
\begin{center}
\includegraphics[width=0.7\columnwidth,clip=]{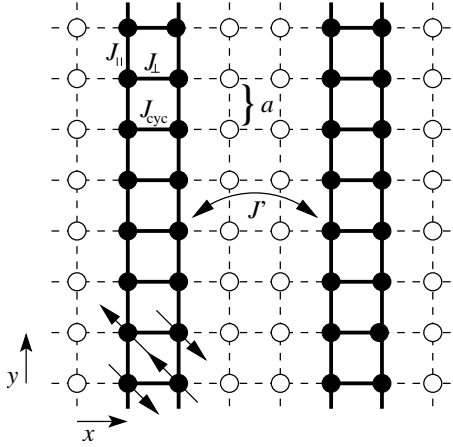}
\end{center}
\caption{Sketch of the model in a single cuprate plane.
Each circle stands for a
copper ion, i.e.\ the site of a spin $1/2$ or a hole. The filled circles
stand for the hole-poor regions where localized spins are assumed. The empty
circles stand for the hole-rich stripes where itinerant behaviour is assumed.
The spin-only model comprises only the localized spins which form a lattice
of coupled two-leg ladders. The intraladder couplings are $J_\parallel,
J_\perp$ and $J_{\rm cyc}$; the interladder coupling $J'$ is an effective
coupling across the charged hole-rich stripes.}
\label{fig:sketch}
\end{figure}

\paragraph{Model}
In the Introduction we have motivated to consider the single-layer model
sketched in Fig.\ \ref{fig:sketch}.
The Hamiltonian $H$ is conveniently split into an intra-ladder part
$H_{\rm ladder}$ and an inter-ladder part $H_{\rm inter}$
\begin{eqnarray}
\nonumber
H_{\rm ladder} \!\!\!\! &=& \!\!\!\! \sum_{i\in \Gamma} J_\perp
 {\bf S}_i^{\rm L}\cdot {\bf S}_i^{\rm R}
+J_\parallel
({\bf S}_i^{\rm L}\cdot {\bf S}_{i+\delta_y}^{\rm L}
+ {\bf S}_i^{\rm R}\cdot {\bf S}_{i+\delta_y}^{\rm R})
\\
&& \hspace{-1.8cm} +\
J_{\rm cyc}  \sum_{i\in \Gamma} [({\bf S}_{i}^{\rm L}\cdot{\bf S}_{i}^{\rm R})
({\bf S}_{i+\delta_y}^{\rm L}\cdot{\bf S}_{i+\delta_y}^{\rm R})  \\
&& \hspace{-1.8cm} +\
({\bf S}_{i}^{\rm L}\cdot{\bf S}_{i+\delta_y}^{\rm L})
({\bf S}_{i}^{\rm R}\cdot{\bf S}_{i+\delta_y}^{\rm R}) -
 ({\bf S}_{i}^{\rm L}\cdot{\bf S}_{i+\delta_y}^{\rm R})
({\bf S}_{i+\delta_y}^{\rm L}\cdot{\bf S}_{i}^{\rm R})
] \nonumber
\end{eqnarray}
where the superscripts L and R stand for the left and the
right spin on a rung, respectively. The subscript $i=(i_x,i_y)$ denotes
the rung by pointing to its center, i.e.\ the mid-point between
the left (L)and the right (R) spin. The possible
values are $\Gamma = a(4\mathbbm{Z},\mathbbm{Z})$. The shift
$\delta_y$ is given by $(0,a)$. The coupling between the spin
ladders reads
\begin{equation}
H_{\rm inter} = J' \sum_{i\in \Gamma}
 {\bf S}_i^{\rm R}\cdot {\bf S}_{i+4\delta_x}^{\rm L}\ ,
\end{equation}
where $\delta_x=(a,0)$.
As in the previous work we consider the isotropic spin ladder
with $J:=J_\perp =J_\parallel$ because the system is derived from
a square lattice. The cyclic exchange ($J_{\rm cyc}$)
is known to be the  dominant correction to the
nearest-neighbour Heisenberg spin exchange
\cite{schmi90,mulle02a,reisc04}. In the square lattice,
its importance has been proposed \cite{loren99c} and
confirmed \cite{colde01b,katan02a}. Similarly, it has
been proposed for two-leg spin ladders \cite{matsu00b} and
could be confirmed by the analysis of two-triplet bound states
\cite{windt01,nunne02}.
We will use the thus established value $J_{\rm cyc} = 0.2 J_\perp$.
Taking $J_{\rm cyc}$ into account is crucial if one aims at a quantitative
agreement with experimental data \cite{uhrig04a,uhrig05a}.
The effective exchange constant $J'$ across the hole-rich stripes depends on
the state of the  eliminated charge degrees of freedom.
The presence of holes substantially reduces $J'$  relative
to $J$ \cite{uhrig04a}.

In practice, we will fit $J'$ to the experimental data; it takes a small
ferromagnetic value ($J'<0$) of a few percent of $J$ \cite{uhrig04a,uhrig05a}.

\begin{figure}[b]
\begin{center}
\includegraphics[width=0.9\columnwidth,clip=]{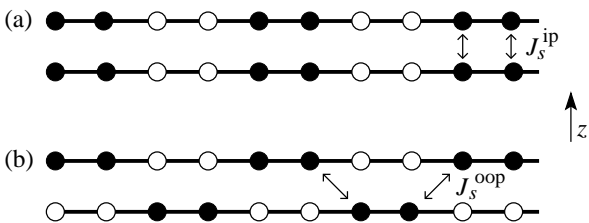}
\end{center}
\caption{Sideview of the two stripe-order patterns
that we consider in bilayer systems (view along the $y$ direction;
open and full circles as in Fig.\ \ref{fig:sketch}).
The pattern in (a) assumes that the modulation is stacked; we
will call this order in-phase. The pattern in (b) assumes that the
modulation is alternating (out-of-phase).}
\label{fig:bilayer}
\end{figure}

Next we extend the above model from a single layer to
a bilayer. We consider the two possibilities
depicted in Fig.\ \ref{fig:bilayer}.
There are, of course, still other patterns which can be considered.
We focus on those in Fig.\ \ref{fig:bilayer} due to their high symmetry.

Denoting the spins in the second layer by the superscripts R' and L'
the coupling between the two layers  reads
\begin{eqnarray}
H_{\rm in-phase} \!\!\!\! &=&\!\!\!\!
J_s^{\rm ip} \sum_{i\in \Gamma}
({\bf S}_i^{\rm L}\cdot {\bf S}_{i}^{\rm L'}
+ {\bf S}_i^{\rm R}\cdot {\bf S}_{i}^{\rm R'})\\
\label{eq:ham_oop}
H_{\rm out-of-phase}\!\!\!\! &=& \!\!\!\!
J_s^{\rm oop} \sum_{i\in \Gamma}
({\bf S}_i^{\rm L}\cdot {\bf S}_{i-2\delta_x}^{\rm R'}
+ {\bf S}_i^{\rm R}\cdot {\bf S}_{i+2\delta_x}^{\rm L'}) .\quad
\end{eqnarray}
The exchange coupling $J_s^{\rm ip}$ takes small antiferromagnetic
values  ($J_s^{\rm ip}>0$).
It corresponds to the interlayer exchange $J_s$ determined for the
undoped parent compound YBCO$_6$ \cite{rezni96,hayde96} so that it
should have about the same value.
In case of YBCO$_6$, a value of $J_s \approx 0.08 J$ has been
deduced from the observation of optical magnons at around 70 meV
\cite{rezni96,hayde96} and from a comparison of the two-magnon spectra in
Raman scattering and in the optical conductivity \cite{gruni00}.

In the coupling $J_s^{\rm oop}$ we do not denote
the small superexchange processes to next-nearest neighbours.
The coupling $J_s^{\rm oop}$ is  an effective
coupling which takes processes into account via the
eliminated site (open circles in Fig.\ \ref{fig:bilayer}b).
If there is a spin on the eliminated site there will be a
ferromagnetic contribution to $J_s^{\rm oop}$ because
both adjacent spins prefer to be antiparallel to it, hence parallel
to each other. If the eliminated site is empty there will be an
antiferromagnetic contribution to $J_s^{\rm oop}$ because
the adjacent spins can undergo exchange processes.
In total, the ferromagnetic and the antiferromagnetic contributions will
partially cancel so that a small  $|J_s^{\rm oop}| <  J_s^{\rm ip}$
remains. Its sign is plausibly ferromagnetic since the eliminated
site is occupied to 75\% at doping $1/8$.
Note that the effective coupling $J'$ between adjacent ladders
was motivated in the same way \cite{uhrig04a}.

\paragraph{Calculation: Single Layer}
The calculation runs in analogy to the one that we
performed previously for the single layer \cite{uhrig04a}.
A perturbative continuous unitary transformation (CUT) is performed
for each ladder separately \cite{knett01b,schmi03d}. Thereby, we
obtain an effective model in terms of the elementary triplet
excitations, triplons \cite{schmi03c}, of the spin ladder.
This effective model includes the one-triplon part
$\sum_{k,\alpha} \omega^0_k \; t^{\alpha,\dagger}_k \; t^\alpha_k\ $
where $k$ is the wave vector along the ladder
and $\alpha\in\{x,y,z\}$
is the flavour index for the three states of a triplon.
The operator $t^\alpha_{k}$ stands for the annihilation of a triplon of
flavour $\alpha$ with momentum $k$
and $t^{\alpha,\dagger}_k$ stands for its creation.
The dispersion $\omega^0_k$ has been computed in many
ways, see e.g.\ Refs.\
\cite{barne93,barne94,trebs00,knett01b,nunne02} or
Fig.\ 2a in Ref.\ \cite{uhrig04a}.

Besides the one-triplon part the effective model  for the
isolated ladder includes contributions involving two and more triplons.
We neglect them since the one-triplon contribution
dominates for the physically relevant parameters
where 63\% of the total spectral weight is in the one-triplon channel
\cite{uhrig04a}.

The coupling \emph{between} the ladders
can be written without a new comprehensive calculation. We apply the
above unitary transformation $U$  to each spin component at each site of a
ladder yielding an effective observable
\begin{eqnarray}
\nonumber
S_{i,{\rm eff}}^{\alpha,{\rm R}}
&:=& U^\dagger S_{i}^{\alpha,\text{R}} U\\
\label{eq:wright}
&=&
\sum_{m\in\mathbbm{Z}} a_m \, (t^{\alpha,\dagger}_{i+m \delta_y}
+ t^\alpha_{i+m \delta_y})
+ \ldots\\
\label{eq:wleft}
S_{i,{\rm eff}}^{\alpha,{\rm L}}
&=&
-\sum_{m\in\mathbbm{Z}} a_m \, (t^{\alpha,\dagger}_{i+m \delta_y}
+ t^\alpha_{i+m \delta_y})
+ \ldots \ .
\end{eqnarray}
Here $t^\alpha_{i}$ stands for the annihilation of a triplon of
flavour $\alpha$ on the dimer $i$ and $t^{\alpha,\dagger}_i$
stands for its creation. Consistent with the above approximation,
the higher contributions involving two and more triplons
are neglected.

First, we will use the identities (\ref{eq:wright}) and (\ref{eq:wleft})
to find the transformed Hamiltonian for the interladder coupling.
This can be done replacing the product of spin components
$S^\alpha_i S^\alpha_{i+4\delta_x}$ by the right hand sides
of Eqs.\ (\ref{eq:wright}) and (\ref{eq:wleft})
taking into account
that triplon operators on different ladders are involved. They
are distinguished in real space by the $x$-component of $i$, i.e.\
$i_x$. In reciprocal space we use $h$ for the momentum perpendicular
to the ladders ($k$ is the momentum along the ladders). This yields
\begin{equation}
H_{\rm inter} = -J'\sum_{h,k;\alpha}d_{h,k}
(t^{\alpha,\dagger}_{h,k}+t^{\alpha}_{-h,-k})(t^{\alpha}_{h,k}+
t^{\alpha,\dagger}_{-h,-k})
\end{equation}
where the minus sign results from the coupling of a
right spin (on rung $i$) to a left spin (on $i+4\delta_x$).
The momentum dependence is given by
\begin{equation}
\label{eq:dhk}
d_{h,k} := \cos(8\pi h) \, a^2(k)
\end{equation}
where $h$ is measured in rlu; the cosine term captures the
shift $4a$ from one ladder to the next one.
The factor $a(k)$ is the Fourier transform of the coefficients $a_m$
\begin{equation}
a(k) = \sum_m \exp({\rm i}2\pi k m)a_m \ .
\end{equation}
The shape of the weight $a^2(k)$ has been illustrated in Fig.\ 2b
in Ref.\ \cite{uhrig04a}.

The total Hamiltonian after the CUT and after the neglect of
the multi-triplon parts is
\begin{eqnarray}
\label{eq:hamilton}
H_{\rm ladder} + H_{\rm inter} &=&
\\
\nonumber
&&\hspace{-3.4cm}
\sum_{h,k;\alpha} \omega^0_k \; t^{\alpha,\dagger}_k \; t^\alpha_k
-J' d_{h,k}
(t^{\alpha,\dagger}_{h,k}+t^{\alpha}_{-h,-k})(t^{\alpha}_{h,k}+
t^{\alpha,\dagger}_{-h,-k}) \ .
\end{eqnarray}
It is bilinear in the triplon creation and annihilation operators
so that it looks like a trivial one-particle problem.
Unfortunately, this is not the case because the triplons are
hard-core bosons. At maximum one of the three kinds may be present
on a given rung.
To make analytical progress, we exploit the fact
that $J'$ is small compared to the global energy scale $J$. Hence the
off-diagonal terms such as $t^{\alpha}_{-h,-k}\;t^{\alpha}_{h,k}$
and its hermitean conjugate
are small and so will be the errors that are introduced by neglecting
the hard-core constraint. If the hard-core constraint is neglected,
that means the triplons are treated as ordinary bosons, the
Hamiltonian (\ref{eq:hamilton}) can be diagonalized by a standard
Bogoliubov transformation.
The resulting dispersion reads
\begin{equation}
\omega_{h,k}  =  \sqrt{(\omega^0_k)^2-4J'd_{h,k}\omega^0_k} \ .
\label{eq:dispersion}
\end{equation}
Note that the omission of the hard-core constraint implies
that there is only one mode per each momentum.
The decay into two or more modes has been neglected.
Thus we are dealing
with a single-mode approximation.

At zero temperature, the dynamic structure factor
$S_{h,k}(\omega)$ measures at which rate
the system can be excited at a given momentum $h,k$
and frequency $\omega$. The excitation operator is the Fourier transform
of the local spin components
\begin{equation}
S^\alpha_{h,k} = \frac{1}{\sqrt{N}}
\sum_i e^{{\rm i} 2\pi (ki_x+h i_y)}
\left( S_{i}^{\alpha,\text{R}}
e^{{\rm i}\pi h} +
S_{i}^{\alpha,\text{L}}
e^{-{\rm i}\pi h} \right)
\end{equation}
where we assume that the position of a rung
is given by its center. Inserting Eqs.\
(\ref{eq:wright}) and (\ref{eq:wleft}) leads to
\begin{equation}
S^\alpha_{h,k} = {\rm i} \sin(\pi h) a(k)
(t^{\alpha,\dagger}_{h,k}+t^{\alpha}_{-h,-k})\ .
\end{equation}
The dynamic structure factor is the sum over all absolute
excitation amplitudes squared, that means the square
of the absolute values of the
prefactors of the creation operators \emph{after}
the Bogoliubov diagonalization. The evaluation brings us to
\begin{equation}
S_{h,k}(\omega) =\sin^2(\pi h) \, a^2(k)
\frac{\omega^0_k}{\omega_{h,k}}\delta(\omega-\omega_{h,k}) \ .
\end{equation}
If we add the contributions at negative frequencies
proportional to $-\delta(\omega+\omega_{h,k})$ we may rewrite
the expression as the imaginary part of a retarded response
function
\begin{equation}
\label{eq:dyn-struc}
S_{h,k}(\omega) =
-\frac{2}{\pi} \, {\rm Im}\frac{\sin^2(\pi h) \, a^2(k) \, \omega_k^0}
{(\omega+i0_+)^2-\omega^2_{h,k}}\ .
\end{equation}
The above formulae describe the single layer case.

\paragraph{Calculation: Bilayer}
Let us consider the in-phase pattern first. There is an obvious
reflection symmetry between the two layers, cf.\
Fig.\ \ref{fig:bilayer}a. Hence the additional degree
of freedom is accounted for most easily by distinguishing even ($\sigma=1$)
and odd triplons ($\sigma=-1$) with respect to the reflection.
We will call this property the parity of the mode.
The additional part of the Hamiltonian reads
\begin{eqnarray}
H_{\rm in-phase} &=&\\
\nonumber
 &&\hspace{-2.4cm}
J_s^{\rm ip} \sum_{h,k,\sigma; \alpha}\sigma a^2(k)
(t^{\alpha,\dagger}_{h,k,\sigma}+
t^{\alpha}_{-h,-k,\sigma})(t^{\alpha}_{h,k,\sigma}+
t^{\alpha,\dagger}_{-h,-k,\sigma})\ .
\end{eqnarray}
We do not write down the remaining part of the Hamiltonian
$H_{\rm ladder}+H_{\rm inter}$ because it is diagonal in $\sigma$;
it is given by Eq.\ \ref{eq:hamilton}
for each value of $\sigma$ separately.
The resulting dispersion depends on the parity $\sigma$,
\begin{equation}
\label{eq:disper-ip}
\omega_{h,k,\sigma}^{\rm ip}
=  \sqrt{(\omega^0_k)^2+4 a^2(k)\omega^0_k
\left(\sigma J_s^{\rm ip}-J'\cos(8\pi h)  \right)} \ .
\end{equation}
The corresponding dynamic structure factor depends on the
parity $\sigma$ only via the dispersion.
So Eq.\ (\ref{eq:dyn-struc}) is still applicable once
$\omega_{h,k}$ is replaced by $\omega_{h,k,\sigma}^{\rm ip}$.

Now we consider the out-of-phase pattern. There is no obvious
reflection symmetry but a combination of reflection and translation
symmetry. Let us assume that all
the spin ladders lie in the same
plane. Then the difference to the single layer calculation is
the presence of the additional ladders located between those
of a single layer. The exchange $J'$ couples ladders at
distance $4\delta_x$; the exchange $J_s^{\rm oop}$ couples spin ladders
at distance $2\delta_x$. Then the whole problem has become
a problem in a single plane. We can stick to the wave vector
$h$ which now is defined in a larger interval ($h\in [-1/4,1/4]$
instead of $h\in [-1/8,1/8]$) because
also rungs with position $i$ in $a (2\mathbbm{Z},\mathbbm{Z})$
carry spins, not only those at $a (4\mathbbm{Z},\mathbbm{Z})$.
The additional contribution in the Hamiltonian  is denoted
\begin{eqnarray}
\label{eq:hamil-oop}
H_{\rm out-of-phase} &=&
-J_s^{\rm oop} \sum_{h,k; \alpha} \cos(4\pi h) a^2(k)\cdot\\
\nonumber
 &&\hspace{0cm}
\cdot (t^{\alpha,\dagger}_{h,k}+ t^{\alpha}_{-h,-k})(t^{\alpha}_{h,k}+
t^{\alpha,\dagger}_{-h,-k})\ .
\end{eqnarray}
where the cosine factor has half the argument of the cosine factor
in Eq.\ (\ref{eq:dhk}) because the distance bridged is only half the one
bridged by $J'$.
The Bogoliubov diagonalization leads to the dispersion
\begin{equation}
\label{eq:disper-oop}
\omega_{h,k}  \!= \! \sqrt{(\omega^0_k)^2-4 a^2(k)\omega^0_k
\left(J_s^{\rm oop}\cos(4\pi h)+J'\cos(8\pi h)  \right)} \, .
\end{equation}

Equations (\ref{eq:hamil-oop}) and (\ref{eq:disper-oop})
start from the doubled Brillouin-zone interval
for the transverse momentum $h$.
In order to compare the in-phase and the out-of-phase patterns
as closely as possible we prefer to stay with the Brillouin-zone interval
of the in-phase pattern. The part of the branch of the out-of-phase pattern
which is located at $|h|\in [1/8,1/4]$ is folded back by the shift
$h= h' \pm 1/4$. The backfolded branch
can be identified with the \emph{odd}
one because the phase factor from a ladder in the lower
layer to a ladder in the upper layer is
$\exp(4\pi {\rm i} h)= -\exp(4\pi {\rm i} h')$,
that means there is an extra factor $-1$ as it
is characteristic for the odd mode. In the dispersion,
we distinguish the two branches by $\sigma =\pm 1$
\begin{eqnarray}
\label{eq:disper-oop2}
\omega_{h,k,\sigma}^{\rm oop}  &=&\\
\nonumber
&&\hspace{-1.8cm}
  \sqrt{(\omega^0_k)^2-4 a^2(k)\omega^0_k
\left(\sigma J_s^{\rm oop}\cos(4\pi h)+J'\cos(8\pi h)  \right)} \ .
\end{eqnarray}
Note that we simplified the notation by passing  from $h'$ back to $h$.
The corresponding dynamic structure factor is again the same
as in the single layer calculation (\ref{eq:dyn-struc}) except that
$\omega_{h,k}$ is replaced by $\omega_{h,k,\sigma}^{\rm oop}$.
Note that the substitution $h\to h \pm 1/4$
does not apply to the sine-factor in the structure factor (\ref{eq:dyn-struc}).
This is so  because the matrix element to excite a local triplet depends
only on the distance between the two spins
on one rung. The phase factor
between the layers does not matter.

Finally, we assume that regions with stripes running vertically and regions
with stripes running horizontally are equally present in the samples.
Hence we will display and discuss the symmetrized
data $\tilde{S}_{h,k}(\omega) \! = (S_{h,k}(\omega) + S_{k,h}(\omega))/2$.

\section{Single layer}
\label{sec:monolayer}
Results for single layers are presented in our previous work
\cite{uhrig04a,uhrig05a}. Very good agreement was obtained for
realistic values of the coupling parameters. In particular, it was
shown that significant cyclic exchange ($J_{\rm cyc}/J \approx 0.2$)
is needed to reconcile the resonance energy and the global energy scale.
In the framework of the stripe model the resonance appears as the
saddle point in a strongly anisotropic dispersion \cite{vojta04,uhrig04a}.
The dispersion is very large along the ladders, i.e.\ along the stripes,
while it is small perpendicular to them.
The global energy scale is set by the maximum of the dispersion.
It can be determined from the analysis of the dispersion itself
\cite{uhrig05a} or from the analysis of the
momentum-integrated structure factor \cite{uhrig04a}.

For LBCO at 1/8 doping, the global energy scale is $J=127$ meV
and the interladder coupling is $J'=-0.072J$. This value
corresponds to the quantum critical point where the
spin-liquid gap just vanishes.
The actual value will be a little larger because there is
evidence for (weak) long-range magnetic order \cite{fujit04}.

Using the single-layer model, the analysis for underdoped YBCO$_{6.6}$ yields
$J=114$ meV and the interladder coupling $J'=-0.035J$ \cite{uhrig05a}.
The significantly smaller value of $J'$ is implied by the presence
of a finite spin gap.

\begin{figure}[t]
\begin{center}
\includegraphics[angle=270,width=0.8\columnwidth,clip=]
{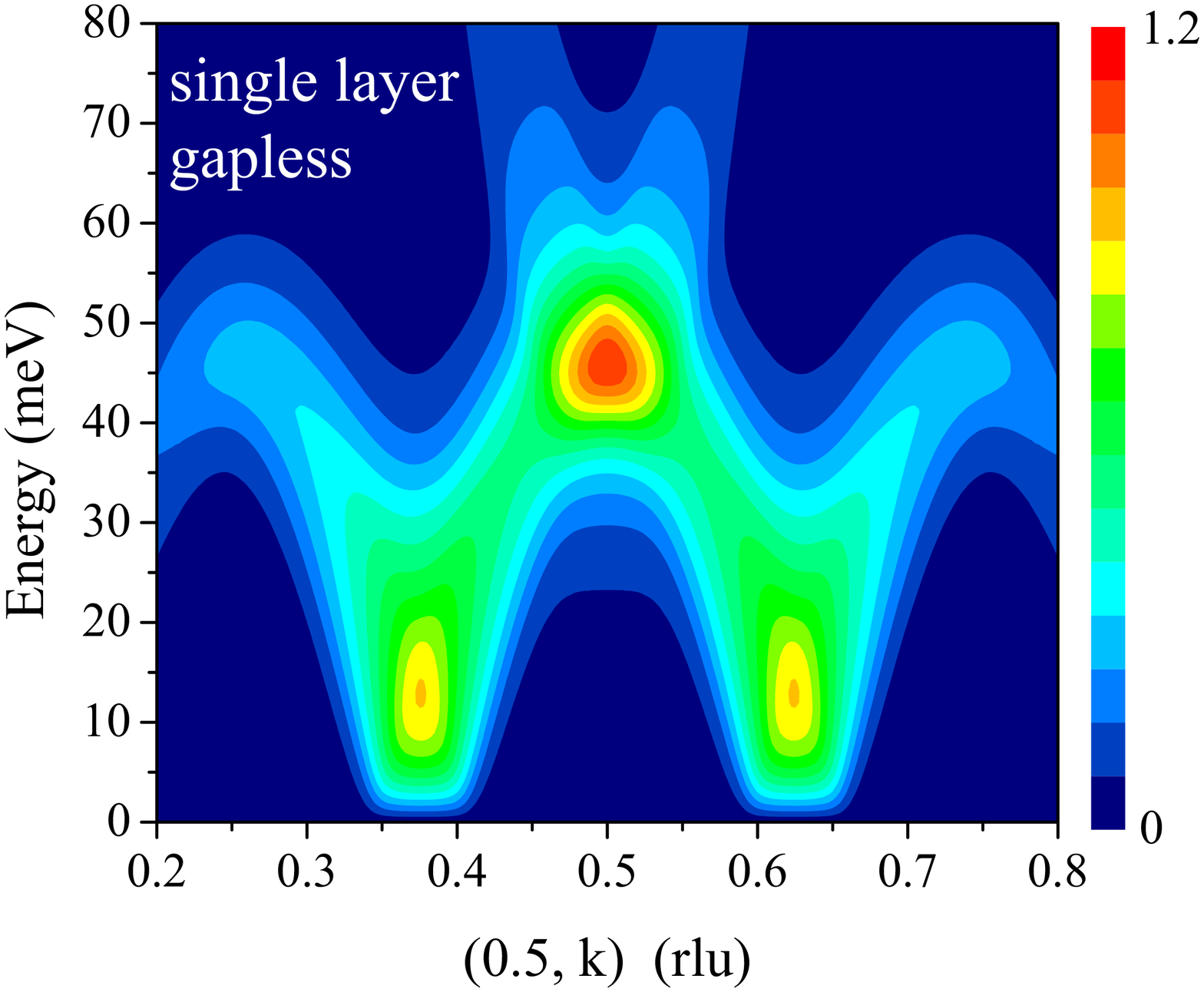}
\end{center}
\caption{Plot of the dynamic structure factor $\tilde{S}_{h,k}(\omega)$
in arbitrary units as function of momentum and energy.
Note that $\tilde{S}_{h,k}(\omega)$ corresponds to the sum over regions with
ladders running in $x$ direction and regions with ladders running in $y$ direction.
The parameters apply to LBCO
at doping 1/8 \cite{tranq04a,uhrig04a},
namely $J=127$ meV,$J_{\rm cyc}/J = 0.2$
and  $J'=-0.072J$. The energy resolution is 6 meV; it is implemented as the
imaginary part $\eta$ in $\omega\to \omega+i\eta$ in expression
(\ref{eq:dyn-struc}). The momentum resolution is $\Delta k = 0.03$ rlu
which means that at a given momentum $(h,k)$
the weight in the square $(h\pm\Delta k,k\pm\Delta k)$ is averaged.
(Color online, gray scale in print version.) }
\label{fig:sl-gapless}
\end{figure}

A cut through the Brillouin zone is displayed in Fig.\ \ref{fig:sl-gapless}.
The cut runs parallel to the reciprocal axis through the
antiferromagnetic wave vector ${\bf Q}_{\rm AF}=(1/2,1/2)$.
Both the energy dependence and the momentum dependence
can be discerned. The values chosen are those which apply to LBCO;
finite resolutions in energy and in momentum are taken into account.

Around $45$ meV a patch of high intensity at ${\bf Q}_{\rm AF}$
is clearly visible. To lower energy the intensity decreases rapidly,
becoming significant again at low energies of 10-20 meV.
There is finite intensity down to the lowest energies since
the system is gapless.
It is very remarkable that the finite resolutions lead to the impression
of almost vertical rods of high intensity (dark (online: yellow) patches)
at the incommensurate positions
$(1/2,1/2\pm 1/8)$. This coincides nicely with the observations of many
experiments, see e.g.\ the generic graph discussed in
Fig.\ 13b in Ref.\ \cite{brom03}. We emphasize that the underlying
dispersion is sine-shaped as expected, see e.g.\ Fig.\ 1 in
Ref.\ \cite{uhrig05a}.

\begin{figure}[t]
\begin{center}
\includegraphics[angle=270,width=0.8\columnwidth,clip=]
{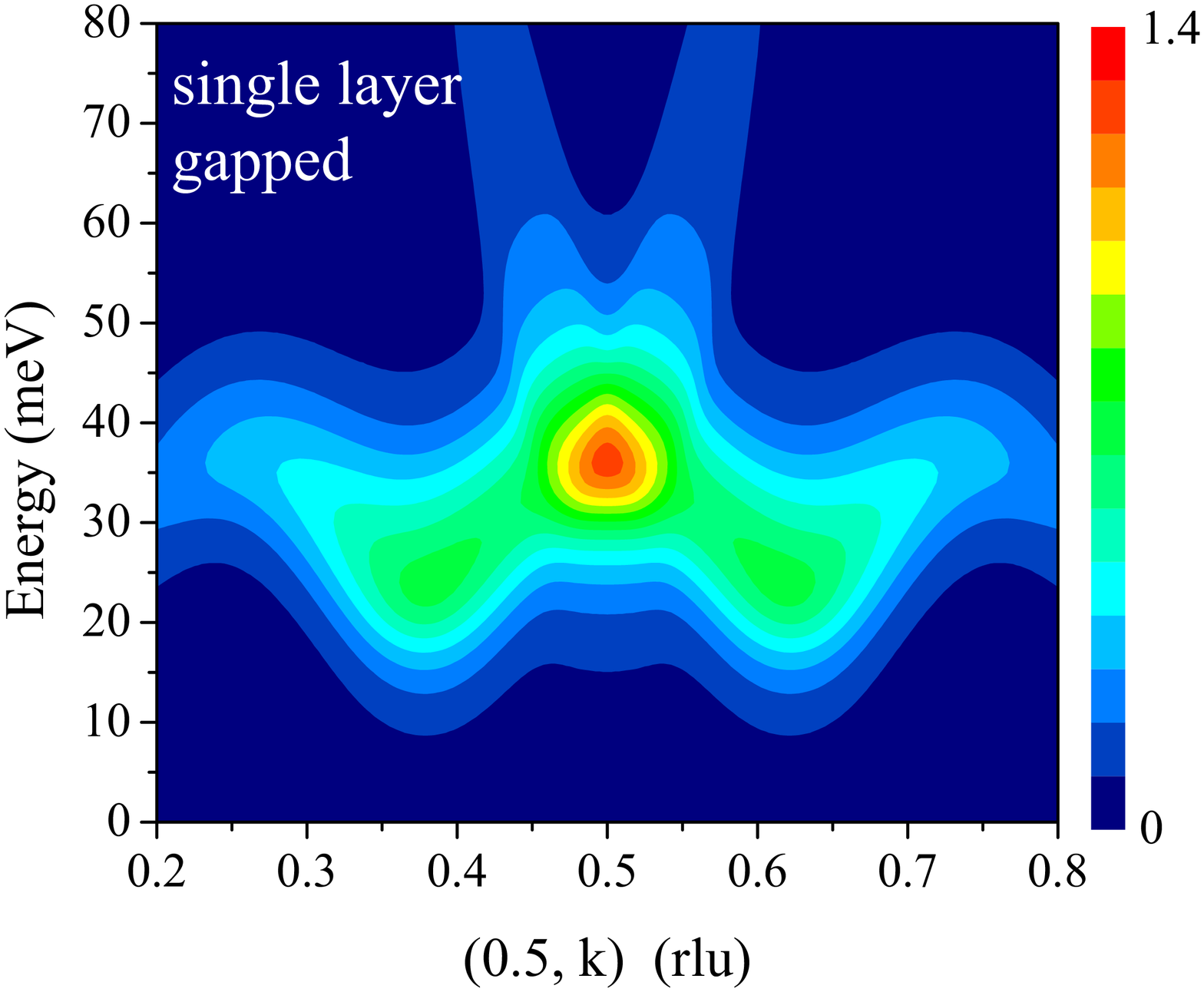}
\end{center}
\caption{Plot of the dynamic structure factor $\tilde{S}_{h,k}(\omega)$
in arbitrary units as function of momentum and energy.
Based on the single-layer model, the parameters
apply to underdoped YBCO$_{6.6}$ \cite{hayde04,uhrig05a},
namely $J=114$ meV,$J_{\rm cyc}/J  = 0.2$
and  $J'=-0.035J$. The plot is to be compared with Fig.\ 1i in
Ref.\ \cite{hayde04}.
The resolutions are chosen as for Fig.\ \ref{fig:sl-gapless}.
(Color online, gray scale in print version.) }
\label{fig:sl-gapped}
\end{figure}

In  Fig.\ \ref{fig:sl-gapped} the corresponding graph is shown for
a gapped system with parameters which apply
to underdoped YBCO$_{6.6}$ (Ref.\ \cite{hayde04})
on the basis of the single-layer model.
The theoretical result agrees well with Fig.\ 1i in Ref.\ \cite{hayde04}{}.
Clearly, the resonance is dominating and there
are tails of intensity to lower energy which point to incommensurate
momenta. The intensity toward lower energies is quickly decreasing
both in theory and experiment.
Discrepancies are seen in the width of the resonance patch
which is larger in experiment. It appears also that the shift
away from ${\bf Q}_{\rm AF}$ is experimentally lower than in our
calculation. Since the experimental result is obtained via
the subtraction of a background it cannot be excluded that finer features
are lost in experiment. Measurements with improved resolution and
signal strength will be very interesting to elucidate further details.

\section{Bilayer}
\label{sec:bilayer}
Now we turn to the properties of bilayer compounds such as YBCO.\@
Before we present
results some considerations on the two patterns under study are in order.
It is plausible that the out-of-phase pattern is energetically
favoured because the charges (the holes in the stripes) are more
evenly distributed, see Fig.\ \ref{fig:bilayer}. An even distribution
lowers the long-range Coulomb potential.
Evidence for such interlayer correlations has been reported
on the basis of a hard x-ray diffraction
study of Nd-doped LSCO \cite{zimme98}. There, stripes in
next-nearest-neighbour  layers are found in an out-of-phase arrangement.
Note that in this particular case, the nearest-neighbour
layers are probably decoupled because the stripes on adjacent layers
are running in orthogonal directions due to the coupling to the lattice via
the tilting of the oxygen octahedra.

\begin{figure}[t]
\begin{center}
\includegraphics[width=0.8\columnwidth,clip=]{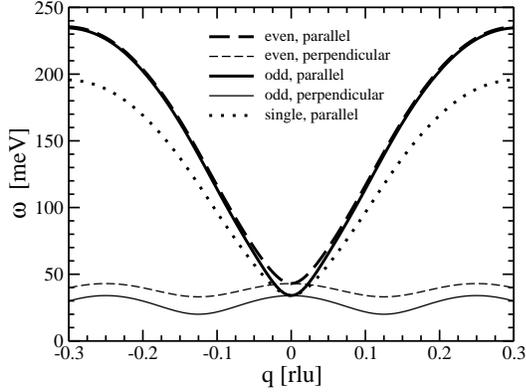}
\end{center}
\caption{Dispersion of the odd (solid) and the even (dashed)
mode for the in-phase pattern as function of $q$, the distance
to the antiferromagnetic wave vector ${\bf Q}_{\rm AF}$.
The thin lines show the dispersion perpendicular to the spin ladders;
the thick lines show the dispersion parallel to the spin  ladders.
The parameters are $J=136$ meV, $J_s^{\rm ip}=0.022J$, $J'=-0.024J$ and
$J_{\rm cyc}=0.2J$, corresponding to $\Delta = 20$ meV,
$\omega_{\rm res}^{\rm even} =43$ meV and
$\omega_{\rm res}^{\rm odd} = 34$ meV.
The dotted line depicts the dispersion in a single layer with
$J=114$ meV, $J'=-0.035J$ and $J_{\rm cyc}=0.2J$, corresponding to
$\Delta = 20$ meV and $\omega_{\rm res} = 34$ meV.
For these parameters, the dispersion perpendicular to the ladders
in the single-layer model (not shown)
is indistinguishable from the dispersion of the odd bilayer mode
(thin solid line).
}
\label{fig:disper-gzs}
\end{figure}

The models for the bilayer have one additional parameter, namely
the interlayer coupling.
The approximate size of this coupling was discussed above following
Eq.\ (\ref{eq:ham_oop}).
We have determined the set of coupling parameters in the following way.
We fix the cyclic exchange at the established
value of $J_{\rm cyc}/J = 0.2$ (see above and
Refs.\ \cite{colde01b,katan02a,nunne02}).
The remaining three parameters $J$, $J'$ and $J_s^{\rm ip/oop}$
are determined from the experimental values for
the spin gap ($\Delta=20$ meV in YBCO$_{6.6}$, Ref.\ \cite{dai99})
and for the energy of the resonance mode in the odd
($\omega_{\rm res}^{\rm odd}$)
and in the even channel ($\omega_{\rm res}^{\rm even}$).
In both channels, the resonance mode corresponds to a saddle point
in the two-dimensional, anisotropic dispersion.
In underdoped YBCO$_{6.6}$ the odd resonance lies at about
$\omega_{\rm res}^{\rm odd} = 34$ meV\cite{hayde04}.
Unfortunately, the even resonance has not yet been observed
at this doping level.
In a slightly overdoped sample, the resonances have been observed at
$\omega_{\rm res}^{\rm even} =43$ meV and $\omega_{\rm res}^{\rm odd} = 36$ meV
\cite{pailh03}.
Since the value of the odd resonance is very similar to the result for
YBCO$_{6.6}$, we assume that also the even resonance is similar and choose
$\omega_{\rm res}^{\rm even} =43$ meV and
$\omega_{\rm res}^{\rm odd} = 34$ meV.

The resulting set of coupling parameters reads
$J=136$ meV, $J_s^{\rm ip}=0.022J$ and $J'=-0.024J$
for the in-phase pattern and
$J=126$ meV, $J_s^{\rm oop}=-0.026J$ and $J'=-0.040J$
for the out-of-phase pattern.
These two sets will be used in all the plots shown in the following.
First, we note that $J_s^{\rm ip}$ is indeed antiferromagnetic
 and $J_s^{\rm oop}$ is ferromagnetic.
Second, the absolute value of $J_s^{\rm ip}$ is fairly small, much smaller than
the interlayer coupling in undoped YBCO$_6$ of $0.08J$.
This is another indication that the out-of-phase pattern is more realistic.

Note that $J$ and  $J'$ do not change much in the bilayer
analyses with respect to the single layer model.
The parameter sets all
are in an experimentally reasonable range.

In Fig.\ \ref{fig:disper-gzs} the dispersions for a single layer and
for the bilayer with in-phase pattern are shown.
The low-lying dispersion perpendicular to the spin ladders is (almost) the
same for the  odd mode and for the single layer. This fact corroborates
the often made statement that the physics of the odd mode in the bilayer
is well described in a model of a single layer. But this does not hold
for the dispersion parallel to the spin ladders where the
single-layer model implies a significantly lower maximum value.
This results from the larger energy scale $J=136$ meV
(instead of 114 meV) needed to meet the resonance energies and the spin gap.
The calculated dispersion of the odd mode appears to be steeper
than observed experimentally \cite{hayde04}, but this will have to be
clarified by more detailed experiments.

It is striking that the odd and the even mode energies are almost
identical above the resonance energies, i.e.\ above $\approx 50$ meV.\@
This can be attributed to the weight factor $a^2(k)$ in
Eq.\ \ref{eq:disper-ip},
which suppresses the difference between the even and the odd mode for small
momenta $k$ (see Fig.\ 2b in Ref.\ \cite{uhrig04a}).
At low energies, i.e., $k\approx 0.5$ rlu parallel to the ladders,
the coupling between the layers leads to the expected splitting of
the two modes.

\begin{figure}[t]
\begin{center}
\includegraphics[width=0.8\columnwidth,clip=]{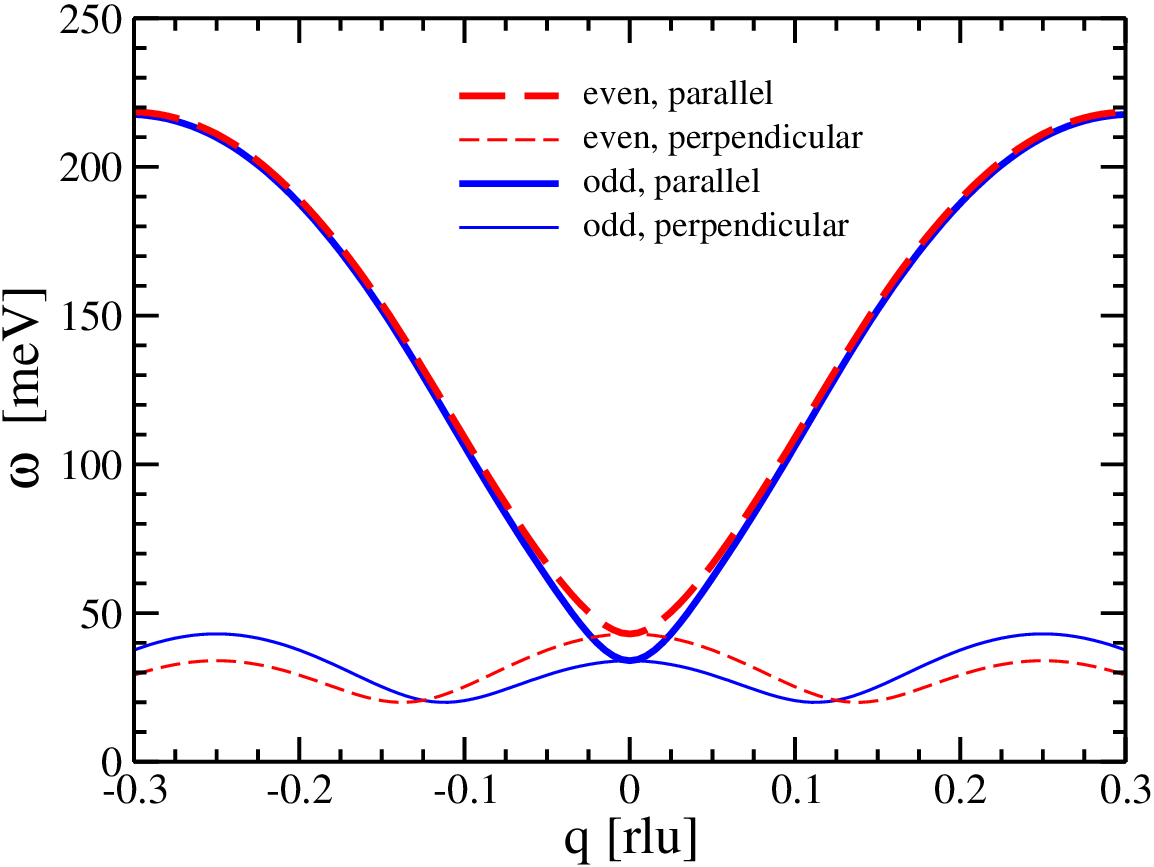}
\end{center}
\caption{Dispersion of the odd (solid) and the even (dashed)
mode for the out-of-phase pattern as function of $q$
as in Fig.\  \ref{fig:disper-gzs}. The parameters are
$J=126$ meV, $J_s^{\rm oop}=-0.026J$, $J'=-0.040J$ and $J_{\rm cyc}=0.2J$.}
\label{fig:disper-azs}
\end{figure}

In Fig.\ \ref{fig:disper-azs}, the dispersions for the
bilayer with out-of-phase pattern are shown.
Again, the odd and the even dispersion almost coincide at energies above the
resonances (saddle points).
The total dispersion is lower than the one for the in-phase pattern
in Fig.\ \ref{fig:disper-gzs} but still higher
than in the single-layer model.

At low energies, the behaviour of the  dispersion perpendicular to
the ladders is remarkable. It is not just a splitting but the dispersion
of one mode can be obtained from the dispersion of the other mode by a
translation by 1/4 rlu perpendicular to the ladders.
This is obvious in view of the derivation of the
dispersions from the common expression (\ref{eq:disper-oop}).
If the dispersion of the even mode can be detected,
our predictions in Figs.\ \ref{fig:disper-gzs} and \ref{fig:disper-azs}
will allow to distinguish both correlation patterns clearly.
The observation of a splitting at $(1/2,1/2\pm 1/8)$
would support the in-phase pattern; the absence of such a splitting
would corroborate the out-of-phase scenario.
Note that in the out-of-phase pattern \emph{both} modes have the
same minimum energy, i.e.\ the same spin gap.

Another very noteworthy observation concerns the position where
the minimum energy is reached. In spite of the underlying
commensurate charge modulation, the spin gap is not reached at
$q=\pm 1/8$ but at a smaller value (in the odd channel).
This results from the sum
of the two cosine-terms under the square root in the dispersion
(\ref{eq:disper-oop2}). Clearly, this effect will be enhanced if
$J_s^{\rm oop}$ increases relative to $J'$.
It opens up the possibility to vary the incommensurability continuously
\emph{without} any change in the periodicity of the charge modulation.
This may give rise to a difference in the saturation value of the
incommensurability between single layer and bilayer compounds.
Such a difference has been discussed in the literature
(LSCO:\cite{yamad98} $\approx 1/8$; YBCO:\cite{dai01} $\approx 1/10$).
Note, however, that the determination of the incommensurability may be
hampered by the asymmetric distribution of the spectral weight (see below).
\begin{figure}[t]
\begin{center}
\includegraphics[angle=0,width=\columnwidth,clip=]{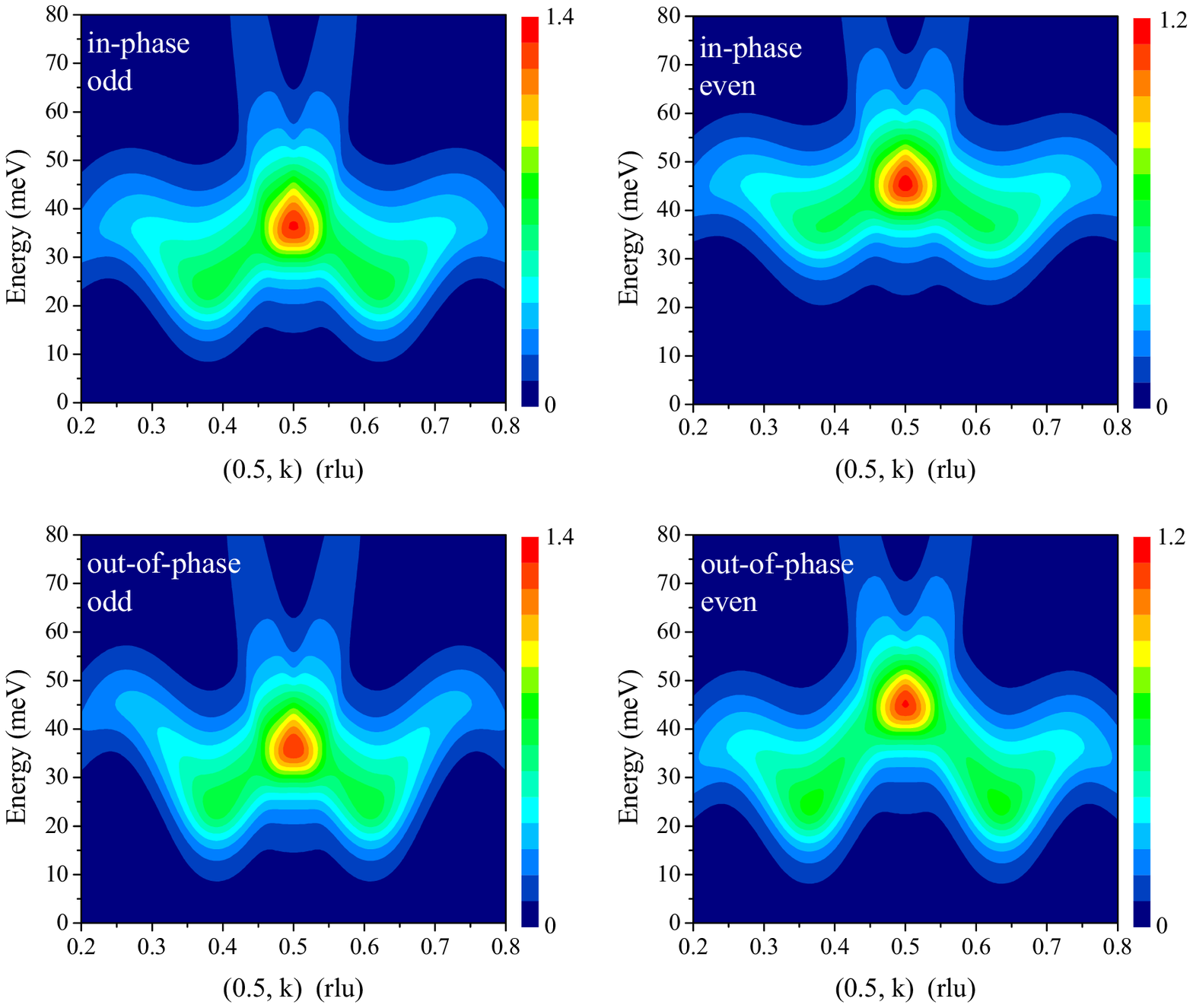}
\end{center}
\caption{Plots of the dynamic structure factor $\tilde{S}_{h,k}(\omega)$
(\ref{eq:dyn-struc})
for the coupling values as in Figs.\ \ref{fig:disper-gzs} and
\ref{fig:disper-azs}. The resolutions in energy and momentum
are those used also in Fig.\ \ref{fig:sl-gapless}.
(Color online, gray scale in print version.) }
\label{fig:bl-cut}
\end{figure}

The Figs.\ \ref{fig:disper-gzs} and \ref{fig:disper-azs} do not provide
information on the intensities. This experimentally crucial information
is displayed in Fig.\ \ref{fig:bl-cut}. The resonance at ${\bf Q}_{\rm AF}$
is dominating in all four plots. The intensity decreases significantly
towards lower energies. The response of the odd  modes is fairly
similar to the response in a single layer (see Fig.\ \ref{fig:sl-gapped}),
in particular the response for the in-phase pattern.
For the out-of-phase pattern, the odd response
displays the interesting feature that the resonance is lower
in energy than the local maxima of the dispersion at $k=1/2\pm 1/4$.

By construction, the resonance of the even modes lies higher in energy.
For the in-phase pattern the whole response is located at higher energies
than in  the odd channel. For the out-of-phase pattern this is not true.
There, the intensity of the even mode reaches down to the spin-gap energy
of the odd channel.
Opposite to the behaviour of the odd dispersion, the local maxima
at $k=1/2\pm 1/4$ in the even channel are located at
lower  energies than the even resonance itself.

The results for the odd mode are certainly
consistent with the available experimental data.
Our spin-only model, however, predicts almost the same intensities for
the even and the odd modes.
This is at odds with experiment where it seems to be extremely difficult
to detect the even mode at all. We presume that the energetically higher
lying even modes are affected more strongly by damping effects, for instance
due to the eliminated charge degrees of freedom and due to the hard-core
triplon-triplon interaction. A strong damping may hinder the experimental
observability significantly. This issue certainly calls for further
theoretical investigations.
However, the damping  should be less important at low energies.
Thus detailed experimental investigations of the spin gap in the even channel
appear very promising.

\begin{figure}[t]
\begin{center}
\includegraphics[width=\columnwidth,clip=]{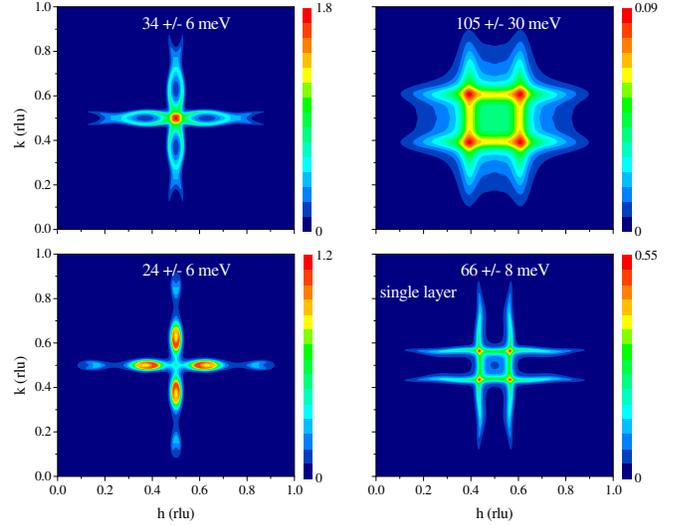}
\end{center}
\caption{Constant-energy scans for the single-layer model. The energies
and the energy resolutions are given in the figure. There is no broadening
in momentum space which must be kept in mind when comparing to experimental
data \cite{hayde04}. The couplings are $J=114$ meV,$J_{\rm cyc}/J  = 0.2$
and  $J'=-0.035J$ \cite{uhrig05a}.
(Color online, gray scale in print version.) }
\label{fig:sl-con-energ}
\end{figure}

\begin{figure}[t]
\begin{center}
\includegraphics[width=\columnwidth,clip=]{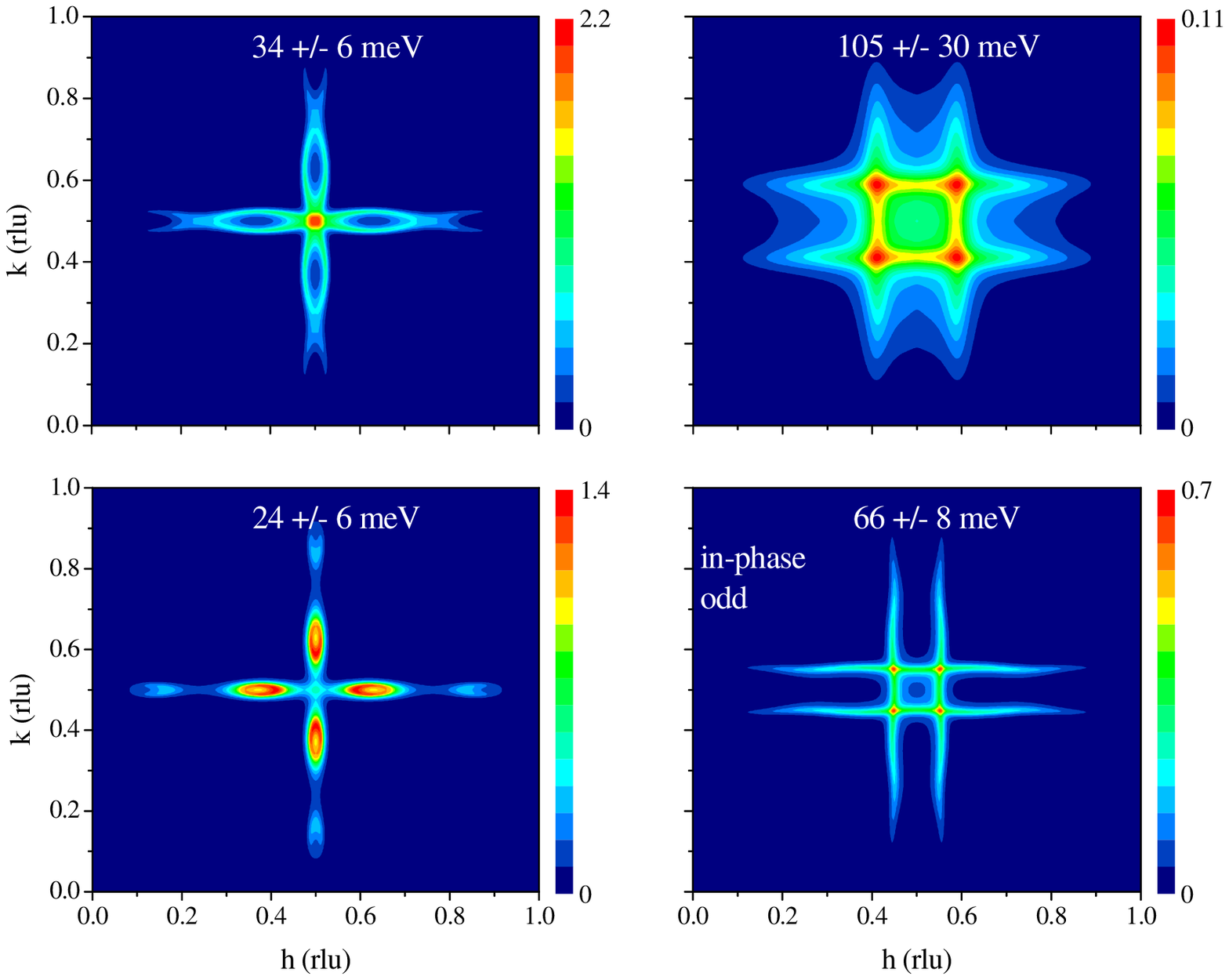}
\end{center}
\caption{Constant-energy scans for the odd mode in the in-phase
bilayer model. The parameters are
$J=136$ meV, $J_s^{\rm ip}=0.022J$, $J'=-0.024J$ and $J_{\rm cyc}=0.2J$.
(Color online, gray scale in print version.) }
\label{fig:odd-ip-con-energ}
\end{figure}
\begin{figure}[t]
\begin{center}
\includegraphics[width=\columnwidth,clip=]{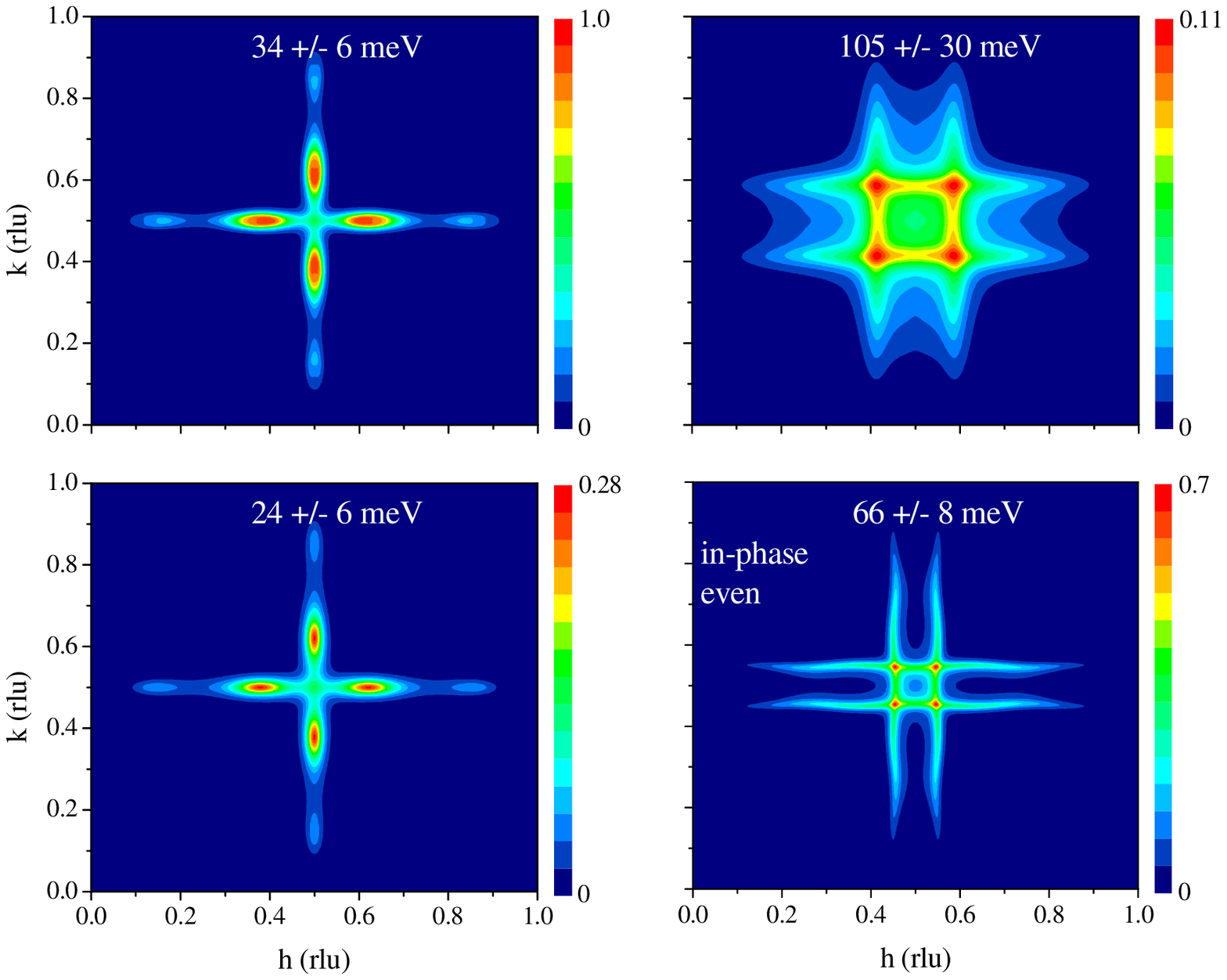}
\end{center}
\caption{Constant-energy scans for the even mode in the in-phase
bilayer model. Parameters are the same
as for Fig.\ \ref{fig:odd-ip-con-energ}.
(Color online, gray scale in print version.) }
\label{fig:even-ip-con-energ}
\end{figure}

\section{Constant-Energy Scans}
\label{sec:scans}
The observations made in the previous section are supported by
constant-energy scans. For comparison we
include in Fig.\ \ref{fig:sl-con-energ} four scans for a single layer.
The energies and the energy resolutions chosen correspond to the
experimental ones \cite{hayde04}. The scan at 24 meV is very close
to the spin gap, i.e.\ the lower bound of the magnetic excitations.
The scan at 34 meV provides data at the resonance energy which
corresponds to the saddle point of the dispersion \cite{uhrig04a,uhrig05a}.
The two remaining scans provide information at energies above the
saddle point.

The theoretical results agree very well with the experimental findings
\cite{hayde04}. In particular, one has a diamond-like shape at low energies
of four incommensurate satellites which merge at the resonance energy to an
almost circular single patch at ${\bf Q}_{\rm AF}$.
Note that already at 24 meV the peak intensity has shifted towards
${\bf Q}_{\rm AF}$, which may blur the precise value of the incommensurability
in the analysis of experimental data.
Above the resonance energy the
shape is square-like with maxima of the intensity at the corners.
The theoretical result at higher energies displays a very low
intensity at  ${\bf Q}_{\rm AF}$ while the experiment shows
squares with non-negligible intensity inside. Furthermore, the tails
of intensity outside the squares are only very vaguely seen in experiment.
We attribute both features to multi-triplon contributions and
life-time effects \cite{uhrig04a} which we neglected in our approach.
But in view of the simplicity of the model, the agreement obtained is
remarkable.

\begin{figure}[t]
\begin{center}
\includegraphics[width=\columnwidth,clip=]{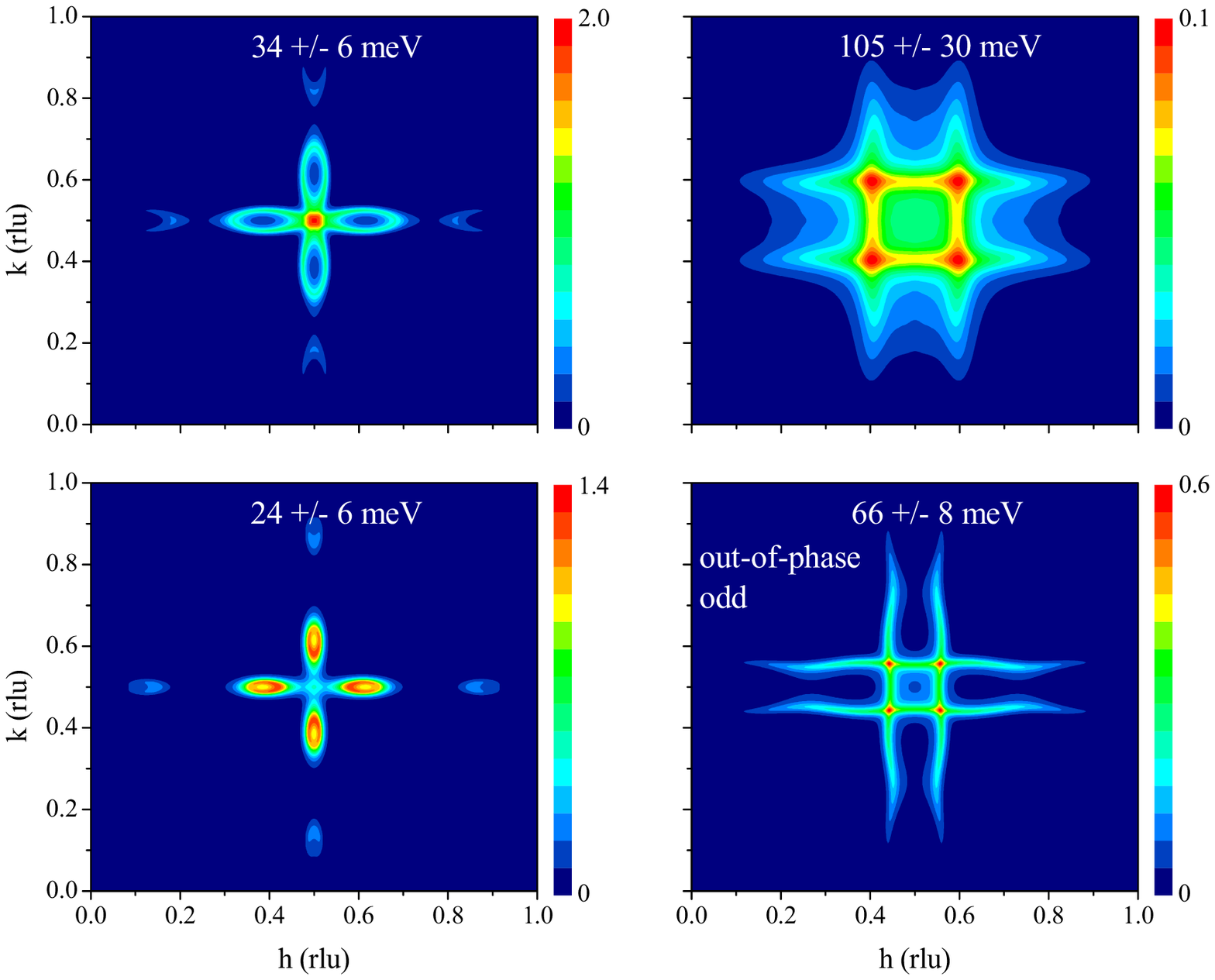}
\end{center}
\caption{Constant-energy scans for the odd mode in the out-of-phase
bilayer model.
The parameters are
$J=126$ meV, $J_s^{\rm oop}=-0.026J$, $J'=-0.040J$ and $J_{\rm cyc}=0.2J$.
(Color online, gray scale in print version.) }
\label{fig:odd-oop-con-energ}
\end{figure}
\begin{figure}[t]
\begin{center}
\includegraphics[width=\columnwidth,clip=]{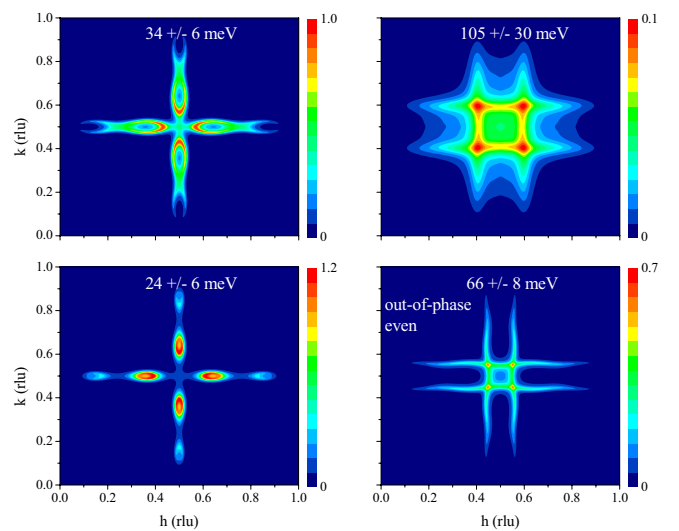}
\end{center}
\caption{Constant-energy scans for the even mode in the
out-of-phase bilayer model. Parameters are the same
as for Fig.\ \ref{fig:odd-oop-con-energ}.
(Color online, gray scale in print version.) }
\label{fig:even-oop-con-energ}
\end{figure}

In Fig.\ \ref{fig:odd-ip-con-energ} the same scans are depicted
for the odd triplon in the in-phase pattern. Qualitatively, these
plots are identical to the ones for the single layer in
Fig.\ \ref{fig:sl-con-energ}. At low energies this is even true
on a quantitative level which can be taken
as an argument that a single-layer calculation
captures a significant portion of the physics
of the odd modes in a bilayer system. At higher energies,  however, the
structures are narrower (i.e.\ closer to ${\bf Q}_{\rm AF}$) in the
bilayer due to the larger value of $J$, as mentioned in the
discussion of the dispersion shown in Fig.\ \ref{fig:disper-gzs}.

In Fig.\ \ref{fig:even-ip-con-energ} the scans for the even mode
are shown. At high energies they are almost identical to those
for the odd mode in  Fig.\ \ref{fig:odd-ip-con-energ}. The most striking
differences occur at low energies where the intensities are
much lower. Note that the scan at 24 meV runs \emph{below} the spin gap of
the even mode, the finite intensity is caused only by the broadening.
The scans for the odd and the even mode at 34 meV differ decisively
because only in the former case the saddle point is hit. This
explains the differences in shape and in maximum intensity.

The results for the out-of-phase pattern are presented in
Figs.\ \ref{fig:odd-oop-con-energ} and \ref{fig:even-oop-con-energ}.
Again, the odd and the even response are almost identical at high
energies while they differ at lower energies. The most striking
difference is whether the resonant saddle point is hit (at 34 meV for the
odd mode) or not (at 34 meV for the even mode). Very interesting is
the situation at 24 meV.
In the in-phase pattern
(see Figs.\ \ref{fig:odd-ip-con-energ} and \ref{fig:even-ip-con-energ})
the even mode
and the odd mode yield the same shape but the intensity differs
by a factor of 5. In the out-of-phase pattern there is no
significant difference in the intensities between the odd and the
even result. But a closer look reveals that the diamond in
the odd channel is smaller than the one in the
even channel (see also the dispersion in Fig.\ \ref{fig:disper-azs}).
This results from the fact that in spite of the given charge
modulation the position of the minima depends on the
ratio of the prefactors of the two cosine-terms in the dispersion
(\ref{eq:disper-oop2}), in particular on the sign of this ratio.
For ferromagnetic $J_s^{\rm oop}<0$ the minima in the odd channel
are closer to ${\bf Q}_{\rm AF}$ than without interlayer coupling
while the minima in the even channel are further away from
${\bf Q}_{\rm AF}$.
These low energy features should be robust against the damping by charge
degrees of freedom and thus appear as promising candidates for future
experiments.

\section{Conclusions}
\label{sec:conclusio}
We investigated a spin-only model with stripe order in a single
layer and for two patterns in a bilayer. In the first pattern,
the stripes and hence the spin ladders are \emph{in-phase} in
the two layers; in the second pattern
they are arranged \emph{out-of-phase}, see Fig.\ \ref{fig:bilayer}.
The patterns are taken to be long-range in our calculation.
But for systems like YBCO we view the assumption of
long-range stripe order as an approximation to fluctuating medium-range
correlations. This approximation significantly eases the theoretical treatment.

The calculation was done by coupling the
effective model, which was obtained previously for isolated
spin ladders with cyclic exchange by continuous
unitary transformations. The effective model is formulated
in terms of triplons, the elementary triplet excitations.
The coupling \emph{between} the ladders is small so that the neglect of the
hard-core constraint of the triplons is justified. Thereby, we
can obtain quantitative results.

The results obtained within the single-layer model
agree very well with experimental
findings, both for gapless \cite{tranq04a,uhrig04a}
and for gapped systems \cite{hayde04,uhrig05a}.
We stress that the coupling parameters $J$ and $J_{\rm cyc}$ are realistic
ones and that only one parameter ($J'$) has not been derived independently
from other experiments. In the gapless case
we have shown that the finite resolution in momentum and energy
leads to the impression of almost vertical rods in the
$(\omega,k)$ plane (Fig.\ \ref{fig:sl-gapless}),
very much like what is seen in many experiments.
In the gapped phase, our calculation
reproduces the experimentally observed rapid decrease
of intensity towards lower energies.

In the bilayer system we distinguish contributions from the
odd and from the even triplon mode. Independent of the
stripe pattern considered the result in the odd channel resembles
very much what is obtained for a single layer. This finding
confirms the often used assumption that a  model with a single layer
suffices to capture the physics in the odd channel of the bilayer
system. The agreement between odd channel and single-layer results
is almost quantitative for the in-phase stripe
correlations, at least at low energies.
For the out-of-phase pattern, the odd mode dispersion does not have
its minimum at $(1/2,1/2\pm \delta)$ with $\delta=1/8$, but
at smaller values $\delta < 1/8$. So we have discovered a way to
reconcile values of $\delta$ different from 1/8 with a charge correlation
dominated by the periodicity $4a$.

The even modes behave almost quantitatively like the odd modes at
energies higher than the resonance energies. At energies of the
order of the resonance energies the even and odd modes split.
In the in-phase pattern this splitting does not depend
on the momentum $h$ perpendicular to the ladders.
In the out-of-phase pattern, however, the even mode dispersion
and the odd mode dispersion can be derived from each other
by a translation by 1/4 rlu perpendicular to the stripes.
This implies that both modes have the same spin gap, that means
the intensity of the even modes reaches also down to low energies.
Furthermore, we predict that the even mode displays a larger
shift in the position of its minima located at
$(1/2,1/2\pm \delta)$ with $\delta>1/8$.
These predictions can be tested experimentally
if the even mode can be observed.

One serious discrepancy between theory and experiment arises from the
relative intensities of the even and the odd modes.
The even mode is hardly seen in experiment. So far it is only seen in the
undoped parent compound YBCO$_6$ \cite{rezni96,hayde96}
and in a slightly overdoped, Ca substituted sample
Y$_{0.9}$Ca$_{0.1}$Ba$_2$Cu$_3$O$_7$ \cite{pailh03}.
In the latter sample
 the intensity is reduced by a factor of 3 with respect to the
odd mode whereas our calculation implies that the reduction in
intensity should only be about 20\%.
This discrepancy can possibly be attributed
to the stronger damping that the
even mode experiences because it lies at higher energy.
A significant damping broadens the response so that it becomes
difficult to distinguish from the omnipresent backgrounds, thus
the signal is lost.
The damping might be due to (i) the eliminated charge degrees of freedom,
(ii) the hard-core interaction between the triplons or
(iii) multi-triplon contributions. The issue which processes are the important
ones must be left to future research.

In this context one may also speculate about a possible doping dependence of
$\omega_{\rm res}^{\rm even}$. Experimentally, it has been established that
the {\em odd} resonance shifts to lower energies with decreasing doping
\cite{sidis04}. Assuming that this shift
is at least partly caused by an increase of the modulus of the
interlayer coupling $J_s^{\rm ip/oop}$,
it corresponds to an {\em increase} of
$\omega_{\rm res}^{\rm even}$ with decreasing doping, and thus also to an
increased damping of the even mode.

We advocate the out-of-phase pattern as the more likely candidate to
describe the relevant bilayer correlations.
First, the concomitant charge distribution is more even, so the Coulomb
interaction favours the  out-of-phase pattern.
Second, the small value of the interlayer coupling of about 2-3\% of $J$
(compared to 8\% in the undoped parent compound) is indicative
that it is not the direct exchange from one layer to the other
as in Fig.\ \ref{fig:bilayer}a which is responsible for
the interlayer coupling. For the out-of-phase pattern
shown in Fig.\ \ref{fig:bilayer}b it is natural to assume that the
(absolute) value of $J_s^{\rm oop}$ is small. Third,
it is appealing to
attribute values of the incommensurability different from 1/8 to
the tunable incommensurability in the out-of-phase pattern.

It is remarkable to which extent the straightforward model
of coupled spin ladders describes the emerging universal
behaviour of the magnetic excitations in high-temperature
superconducting cuprates. This is even more fascinating
in view of the
fact that static stripes are not supported by experiment \cite{hinko04}
and of the tendency
to form more two-dimensional tiling patterns, at least
at the surfaces \cite{hanag04,mcelr05a}.
We think that the universal behaviour of the magnetic excitations
may be explained by assuming
that on a medium-range of 4 to 8 lattice spacings the correlations
are similar to those of stripes.
It is plausible that our results for not too low energies remain
qualitatively valid for such a system with medium-range stripe correlations.
Definitely, more work is called for to
improve the theoretical description further.

Equally, further high-resolution neutron experiments are desirable
in order to test the predictions made in the present work.
This will provide deeper insight in the complex physics of the
elementary excitations of the cuprate superconductors.

\section*{Acknowledgment}
Helpful discussions are acknowledged with M.~Braden, B.~Keimer, S.~Sachdev,
J.M.\ Tranquada, M. Vojta and T. Yokoo.
This work was supported by the DFG in SFB 608
and in SP 1073 as well as by the COE at the Tohoku University, Sendai, where a
part of this work was done during a visiting professorship of one of us (GSU).

%\bibliographystyle{../../bibinput/myjpsj}
%\bibliography{../../bibinput/liter10}

\end{document}